\documentclass[twocolumn,showpacs,amsmath,amssymb, nobibnotes, aps, prl,showkeys]{revtex4}

\usepackage{graphicx}
\usepackage{dcolumn}
\usepackage{bm}
\usepackage{docs}

\expandafter\ifx\csname package@font\endcsname\relax\else
 \expandafter\expandafter
 \expandafter\usepackage
 \expandafter\expandafter
 \expandafter{\csname package@font\endcsname}%
\fi

\begin{document}

\title{Gravitational deflection of light in the Schwarzschild -de Sitter space time}

\author {Arunava Bhadra$^{a}$, Swarnadeep Biswas $^{a,b}$ and Kabita Sarkar $^{c}$}

\affiliation{$^{a}$ High Energy $\&$ Cosmic Ray Research Centre, University of North Bengal, Siliguri, West Bengal, India 734013\\
$^{b}$ Department of Physics, Assam University, Silchar, Assam, India 788011 \\
$^{c}$ GyanJyoti College, Dagapur, Siliguri, West bengal, India 734001 \\
}
\begin{abstract}

Recent studies suggest that the cosmological constant affects the gravitational bending of photons although the orbital equation for light in Schwarzschild-de Sitter space time is free from cosmological constant. Here we argued that the very notion of cosmological constant independency of photon orbit in the Schwarzschild-de Sitter space time is not proper. Consequently the cosmological constant has some clear contribution to the deflection angle of light rays. We stressed the importance of the study of photon trajectories from the reference objects in bending calculation, particularly for asymptotically non-flat space-time. When such an aspect is taken in to consideration the contribution of cosmological constant to the effective bending is found to depend on the distances of the source and the reference objects.

\end{abstract}

\maketitle

\section*{I. Introduction}

A number of recent cosmological observations indicate the presence of the cosmological constant with a value of $\Lambda \simeq 10^{-52} /; m^{-2}$ (e.g. see [1] and references therein). Consequently the exterior space time due to a static spherically symmetric mass distribution is the Schwarzschild-de Sitter (SDS) [2]. Recently working with the SDS geometry, Rindler and Ishak demonstrated [3] that contrary to the widely held idea [4,5] there is a small contribution of the cosmological constant ($\Lambda$) in the gravitational bending of light that diminishes the deflection angle when $\Lambda$ is positive although the orbital equation for light in SDS space time is free from $\Lambda$. In a subsequent work Ishak et al [6] have further shown that the contribution of $\Lambda$ on bending of light could be significant (larger than the second order term) for many lens systems such as cluster of galaxies.  

Some following works [7-9] support the conclusion of Rinder-Ishak. In particular Sereno [7] showed that the deflection angle in SDS space time contains a term that describes local coupling between the lens (characterizes by mass) and $\Lambda$. He provided a general expression for the bending angle and claimed that the results of Rindler and Ishak [3] can be recovered from his expression for a specific radial distance of the observer/source. 

Some investigations [10-12], however, questioned the findings of Rindler-Ishak. The criticisms against Rindler-Ishak method are mainly rest on the fact that in the SDS universe the lens, source and observer are moving relative to each other which has not been incorporated by Rindler and Ishak in their (original) analysis. Khriplovich and Pomeransky [10] accommodated such a dynamic feature by working with Friedmann-Robertson-Walker (FRW) coordinates, whereas Park [11] analyzed the problem by expanding the null geodesic equations followed from McVitte metric to first order in mass in a newly defined physical spatial coordinates consistent with the expanding universe. Both the works concluded that cosmological constant plays no role in gravitational lensing. In order to avoid coordinate dependent artifacts, Simpson et al [12] employed the standard technique of cosmological perturbations and by working in the Newtonian gauge they obtained that the potential in the perturbed FRW metric has no explicit dependence on cosmological constant and thus they concluded that $\Lambda$ dependence of bending angle obtained from the Kottler metric is a gauge artifact result. In a subsequent work Ishak et al [13] addressed the criticisms and pointed out that the conclusion of no contribution of $\Lambda$ to the bending angle is mainly due to improper dropping of relevant terms in calculating the deflection angle. On the other hand Serono [7] argued that the separate $\lambda$ contribution is absent bending expression of [10,11] because it is included in the angular diameter distances. 

Here we question the very concept of $\Lambda$ independency of the orbital equation of light in the SDS geometry. Our reasoning is that the first order differential equation of null geodesic in the SDS geometry contains $\Lambda$ that drops out at the second order. So the solution of the second order differential equation for null geodesic must also satisfy the parent first order differential equation and thereby the orbital solution should include $\Lambda$. When one integrates the second order differential equation of null geodesic, the $\Lambda$ should reappear in the solution through an integration constant. 

More importantly the measurement of bending requires a reference \lq straight \rq line (except in the special case of Einstein ring). Usually path of light rays from a reference source having impact parameter much larger than that for the light rays from the source or the same source but in different position is considered as the reference line for measuring the bending.  But in the SDS geometry light trajectory of such a reference source cannot be treated as \lq straight line \rq. This is because in contrast to the gravitational effect due to mass that falls off quickly with distance, the influence of $\Lambda$ increases with distance from the source. Hence it is expected that in the SDS geometry the light rays from the reference source will also be affected, possibly to a higher degree, by $\Lambda$. 

In the present work we would first compute the bending angle in the SDS space time taking the proper $\Lambda$ involved solution of the light trajectory and then we would show that when the light path from reference source is taken into consideration the resultant bending in the SDS geometry will appear to increase rather than decrease due to $\Lambda$ effect. The possibility of detection of $\Lambda$ effect from the bending angle measurement will be discussed.    

In deriving the deflection angle in the SDS geometry we would assume, as usually done, that the whole lensing system consisting of source, reference, lens and observer, is an isolated one; light rays while moving from the source/reference to the observer is not influenced by any other object outside the system. We would restrict our discussion only to the case of local (within galaxy) scale; we are not going to consider the situation involving large distance scale such as lensing due to cluster of galaxies etc. So the cosmological expansion should not have much influence here. Our sole objective is to estimate the bending angle correctly in the presence of cosmological constant and also to explore whether the $\Lambda$ contribution to the bending angle can be detected experimentally in principle. 

The plan of the paper is the following. In the next section we will first obtain the orbit equation for light rays in the SDS geometry and consequently we would derive the expression for gravitational deflection angle. In section 3 we will estimate the expected deviation in image position of the source with respect to a reference source. Finally we will conclude our results in section 4.      

\section*{II. Gravitational deflection in the SDS space time}

For the computation of deflection angle we would follow the procedure described in [14]. We consider the following geometrical configuration for the phenomenon of gravitational bending of light. The light emitted by the distant source S is deviated by the gravitational source (Lens) L and reaches the observer O. The angles are measured with respect to the polar axis which is parallel to the un-deflected ray (in the absence of massive object) and passes through the center of the lens (L). Such a choice of the polar axis has been justified in [13]. The point L is taken as the origin of the coordinate system. Our first target is to estimate theoretically the deflection angle and subsequently the image position in the context of SDS geometry. 

The metric for the SDS or Kotler space-time is given by (we are using units such that $G=c=1$)

\begin{equation}
ds^{2} = - f_{\Lambda}(r)dt^{2} + \frac{dr^{2}}{f_{\Lambda}(r)} + r^{2}(d\theta^{2}+sin^{2} \theta d\phi^{2} )   
\end{equation}

where
\begin{equation}
f_{\Lambda}(r)=\left( 1-\frac{2m}{r} - \frac{\Lambda r^{2}}{3} \right) \; ,
\end{equation}

$m$ being the mass of the lens obsject. For this space time the null geodesic equation involving $r$ and $\phi$ is given by (see [14])

\begin{equation}
\frac{1}{r^{4}}\left( \frac{dr}{d\phi }\right) ^{2} + \frac{f_{\Lambda}}{r ^{2}} - \frac{1}{ b^{2}} = 0
\end{equation}

where $b \equiv r^{2} d\phi/dp$ is the first integral of motion that behaves as the impact parameter at large distances, $p$ is an affine parameter along null geodesic. Writing $u=1/r$ and differentiating the above equation with respect to $\phi$, one gets the second order differential equation for null geodesic that does not contain $\Lambda$ as given hereunder

\begin{equation}
\frac{d^{2}u}{d\phi^{2}} + u = 3mu^{2}
\end{equation}

which is the same to that for the Schwarzschild space time. The general solution of the above path equation up to the first order accuracy in $m$ reads as 

\begin{equation}
u = \frac{sin \phi}{R} + \frac{3m}{2R^{2}} \left(1+\frac{1}{3}cos 2\phi \right) 
\end{equation}

The above solution must be a solution of Eq. (3) also which implies

\begin{equation}
\frac{1}{R} - \frac{m}{R^{2}}= \left( \frac{1}{b^{2}} + \frac{\Lambda}{3} \right)^{1/2} \;.
\end{equation} 

The Eq.(5) together with the above relation imply that the orbit equation of light rays in the SDS geometry does contain $\Lambda$. 

The co-ordinate angular velocity is given by  

\begin{equation}
\frac{dr}{d\phi} = - \frac{r^{2}}{R} cos \phi \left(1- \frac{2m}{R}  sin \phi \right)
\end{equation}

which vanishes at the coordinate distance of closest approach ($r_{o}$) that occurs when $\phi = \pi/2$. The parameter $R$ is thus related with $r_{o}$ through the following relation

\begin{equation}
\frac{1}{r_{o}} = \frac{1}{R} + \frac{m}{R^{2}}
\end{equation}

hence the coordinate distance of closest approach (Eq.(8)) also depends on $\Lambda$ through Eqs. (6).  

For asymptotically flat space time such as the Schwarzschild space time the direction of asymptotic light rays is usually evaluated by applying the limit $r \rightarrow \infty$ in the orbit equation and the angle between the two asymptotic directions gives the total deflection angle. However, $r \approx \sqrt{3/\Lambda}$ gives the de Sitter horizon. Hence $r \rightarrow \infty$ does not make any sense in SDS space time. This was one of the main objections raised by Rindler and Ishak [3] against the conventional treatment of calculating bending in SDS space time. As a solution they proposed to consider the angle that the tangent to the light trajectory made with a coordinate direction at a given point which for the general metric (1) is given by 

\begin{equation}
tan(\psi) = rf(r)^{1/2}|d\phi/dr|
\end{equation} 

For the null geodesic the above equation reduces to [14]

\begin{equation}
tan(\psi) = \left[\frac{ f(r_{o})}{ f(r) } \frac{r^{2}}{r_{o}^{2}} -1\right]^{-1/2}
\end{equation} 

which to the leading order in $m$ and $\Lambda$ gives

\begin{equation}
tan(\psi) = \frac{r_{o}}{r} + \frac{m}{r} - \frac{m r_{o}}{r^{2}} - \frac {\Lambda r_{o} r}{6} + \frac {\Lambda r_{o}^{3} }{6 r} 
\end{equation} 

When $r>> r_{o}$ the angles $\phi$ and $\psi$ will be small and consequently one may take $sin(\phi) \rightarrow \phi$ and $tan(\psi) \rightarrow \psi$. Using Eq.(5) we get the angle between the tangent to the light trajectory at point ($r, \phi$) and the polar axis to the leading orders in $m$, $\Lambda$ and $r_{o}/r$ 

\begin{equation}
|\epsilon| = |\psi - \phi| = \frac{2m}{r_{o}}  - \frac{m r_{o}}{r^{2}} 
 - \frac{\Lambda r_{o} r}{6}  + \frac {\Lambda r_{o}^{3} }{6 r}
\end{equation} 
 
At this juncture an important question is that at what coordinate point(s) the angle $\psi$ to be determined. It appears that the selection of the points on the orbit at which the tangents to be drawn for estimation of angles remains somewhat arbitrary in the literature. For instance, in their basic work while obtaining bending angle for the SDS metric Rindler and Ishak [3] used the point $\phi =0$ (corresponding $r$ follows from the orbit equation) purely on the basis of convention whereas in a subsequent work [6] the angle was determined at the boundary of the SDS vacuole. While the former choice is not a proper one (the angular position of the observer cannot be taken as zero in the coordinate system considered here, if it is so taken forcefully the observer distance will become fixed by the distance of closest approach as may be seen from Eq.(5) and (8)), the later choice has limited applicability. Thus a straight forward approach should be to calculate the angles directly at the location of the observer and the source.  

The angle of transmission $\epsilon_{s}$ and reception $\epsilon_{o}$ with respect to the polar axis can be straightway computed from Eq. (12) at the location of the source $\left(d_{LS}, \phi_{s} \right)$ and the observer $\left(d_{OL}, \phi_{o} \right)$ respectively and the total deflection angle thus reads 

\begin{eqnarray}
|\epsilon| &=& \frac{4m}{r_{o}} - 2 m r_{o} \left(\frac{1}{d_{LS}^{2}} + \frac{1}{d_{OL}^{2}} \right)  \nonumber \\
&& - \frac{\Lambda r_{o} }{6} \left( d_{OL} + d_{LS} \right) + \frac{\Lambda r_{o}^{3} }{6} \left( \frac{1}{d_{OL}} + \frac{1}{d_{LS}} \right)
\end{eqnarray}  

The distance of closest approach is a coordinate dependent variable; it is not a measurable one. Identifying the measured radius of an object with  coordinate distance of closest approach works tolerably well only up to the first order level for standard or isotropic coordinate sytem but such an approximation does not work at second or higher order in $m$ [15]. It is proper to express the bending angle in terms of a coordinate independent quantity such as the apparent impact parameter $b$. 

The relaton between $b$ and $r_{min}$ may be obtained from Eq.(3)  

\begin{equation}
\frac{1}{r_{o}} - \frac{m}{r_{o}^{2}}= \frac{1}{b} - \frac{\Lambda b}{6}
\end{equation}

Exploiting the above relation, one finally gets to the leading order in m and $\Lambda$ the total deflection angle in terms of impact parameter  

\begin{eqnarray}
|\epsilon| &=& \frac{4m}{b} - 2 m b \left(\frac{1}{d_{LS}^{2}} + \frac{1}{d_{OL}^{2}} \right) + \frac{2m \Lambda b }{3}
\nonumber \\
&&  - \frac{\Lambda b }{6} \left( d_{OL} + d_{LS} \right) + \frac{\Lambda b^{3} }{6} \left( \frac{1}{d_{OL}} + \frac{1}{d_{LS}} \right)
\end{eqnarray}  

One may note that the $\Lambda$ contribution part in the above expression is not the same to the one obtained in [3]. This is mainly due to use of the modified (proper) orbit equation. Moreover, in [3] the angle $\psi$ has been determined at a different (improper) coordinate point. If we forcefully take $\phi=0$ in Eq.(5), then $1/r = 2m/R^{2}$. On substitution of this r, the third term of the right hand side of Eq.(12) becomes $\Lambda R^{3}/(12m)$ which is what Rindler and Ishak found in their work [3]. The third term of the right hand side of Eq.(15) is the one that Serono qualified as local [7] since it contains m and $\Lambda$ but not positional coordinates of the source/observer. 

It appears from the above expression that even in a solar system observation the $\Lambda$ contribution to the bending can be considerable if the source is at a large distance away from us.

\section*{III. Deviation of image positions between the source and the reference object}

The gravitational bending of light trajectories has been measured experimentally with high precision. At the early stages the bending was measured in solar system experiments by comparing the apparent positions of stars when light trajectories from the stars come close to solar disc but remain visible (normally during a solar eclipse) with their positions half an year earlier when the stars were on the opposite side of the Earth from the Sun and thereby light rays from these sources do not come to close to the Sun on their way to the Earth. In modern high precision measurements of gravitational deflection using interferometric technique, angular positions of stars are measured as a function of time with respect to other sources having larger impact parameters treating the later objects as references. For instance in an effort to test the gravitational theories the change in angle between the quasar $3C279$, which is occulted by Sun in each October, and the quasar $3C273$ from their angular separation of about $9.5^{o}$ has been measured just before and after occultation and the results are found in accordance to the prediction of general relativity to the first order accuracy in $M_{\odot}/R_{\odot}$. 

The deviation of the image position ($\theta$) of the source from its actual (would have been seen by the observer in the absence of the Lens) position ($\beta$) can be obtained from the lens equation which is given by [16]

\begin{equation}
tan \beta = \frac{d_{OL}}{d_{OS}} \frac {sin \theta}{cos (\delta - \theta)} - \frac{d_{OS}-d_{OL}}{d_{OS}} tan (\delta - \theta)  \;,
\end{equation}

where the angles are with respect to the optic axis (the line joining the observer and the lens) and $d_{OS}/cos \beta$ is the distance between the observer and the source. Hence in this scenario $\theta = \psi$. For small angles i.e. when $\theta, \beta,\delta << 1$, the lens equation reduces to 

\begin{equation}
\beta \simeq \theta - \frac{d_{LS}}{d_{OS}} \delta
\end{equation}

As mentioned already experimentally the effect of the lens on photon trajectory is obtained by measuring the bending with respect to the photon trajectory from a second source that may be called as reference source. The distance of closest approach for light path from the reference has to be much larger than that for photon trajectory from the source. 

Thus when angles are small, the angular difference between the images of the source and the reference, as to be revealed to the observer, is

\begin{equation}
\theta^{R} - \theta^{S} = \beta^{R} - \beta^{S} + \left(\frac{d_{LR} \delta^{R}}{d_{OR}}  - \frac{d_{LS} \delta^{S}}{d_{OS}}  \right)
\end{equation}

where superscript $R$ and $S$ denote the reference and the source respectively and $d_{OR}/cos \beta^{R}$ is the distance between the observer and the reference object. As observer changes his/her position, both $\beta$ and $\theta$ of the source as well as of the reference will change. The difference between the impact parameter or the closest approach of the light path as the observer changes his/her position from one point to other is expected to be the same for both source and reference. In other words $b_{2}^{S} - b_{1}^{S}$, where the subscript 1 and 2 refer to parameter b at position 1 and 2 of the observer respectively, should be the same to $b_{2}^{R} - b_{1}^{R}$, particularly when the source and reference are at a large distance away from the lens. To the leading order the difference in angle as observer changes position is finally turned out as

\begin{eqnarray}
\delta \alpha &=& \theta^{R}_{2} - \theta^{S}_{2}- (\theta^{R}_{1} - \theta^{S}_{1})    \nonumber \\ 
&& \simeq - 4m \left(\frac{d_{LR}}{d_{OR} b_{1}^{R}} - \frac{d_{LS}}{d_{OS} b_{1}^{S}} \right) - \frac{2m \Lambda b }{3} \left(\frac{d_{LR}}{d_{OR}} - \frac{d_{LS}}{d_{OS} } \right) \nonumber \\
&& + \frac{\Lambda \delta b d_{OL}}{6} \left(\frac{d_{LR}}{d_{OR}} - \frac{d_{LS}}{d_{OS}}  \right) + \frac{\Lambda \delta b d_{OL}}{6} \left(\frac{d_{LR}^{2}}{d_{OR}} - \frac{d_{LS}^{2}}{d_{OS}}  \right) \;,
\end{eqnarray}   

where $\Delta b \equiv b^{i}_{2}-b^{i}_{1}$, $i$ stands for source/reference. Here we take that the impact parameter (or the distance of closest approach) (for both the source and the reference) is smaller at position 1 than at position 2. When both the reference object and the source are far far away from lens in compare to the lens - observer distance, one may take $d_{LR}/d_{OR} \simeq d_{LS}/d_{OS} \simeq 1$. In such a condition the relative deflection angle becomes

\begin{equation}
\delta \alpha \simeq  \frac{4m}{ b_{1}^{S}} - \frac{\Lambda \delta b }{6} \left(d_{LR} - d_{LS}  \right) \;,
\end{equation}    

So the deflection angle up to the accuracy we considered here still contains a cosmological constant involved term unless $d_{LR} = d_{LS}$.  

\section{IV. Conclusion}

We conclude the followings

1) The cosmological constant does affect the gravitational bending angle which is in accordance with the conclusion of Rindler and Ishak [3]. 

2) To the leading order there are two terms involving cosmological constant in the expression of bending, one of them is purely local in the sense that it does not contain any information about the location of the observer/source. Instead this term describes the coupling between the lens and the cosmological constant as first pointed out by Sereno [7]. Interestingly this term has the same signature to that of the classical expression of general relativistic bending (4m/b) i.e. this term will cause an increase of the bending angle. The other term, which is the dominating one, involves the radial distances of the source and the observer and it bears the repulsive characteristics of the positive cosmological constant. In the case of a light ray grazing the limb of the sun and if the source distance is 10 kpc which is roughly equal to the distance of the sun from the galactic centre, the ratio of $\Lambda$ contribution to the main general relativistic contribution is about $10^{-18}$. Note that in solar system the influence of cosmological constant is known to be maximum in the case of perihelion shift of mercury orbit where the $\Lambda$ contribution is about $10^{-15}$ of the total shift.    

3) While studying gravitational bending in Schwarzschild-de Sitter geometry or in any asymptotic non-flat space time it is also important to study the photon trajectories from reference objects with respect to which the bending will be measured. When such an aspect is taken in to consideration the $\Lambda$ contribution to the effective bending is found to depend on the distances of the source and the reference objects (Eq.(19) or Eq.(20)). In principle the $\Lambda$ effect can be detected from the bending angle measurement choosing suitable source and reference objects.   

4) In the instance of formation of Einstein ring, however, no reference object is needed. In that particular case the ring radius will be smaller than that of the Schwarzschild geometry due to $\Lambda$ contribution as noted in [7,17]. 

5) The effect of $\Lambda$ will be prominent for sources of large distances. In such situation, however, the effect of $\Lambda$ due to cosmological expansion may dominate over the geometric term [18]. So far the contribution of cosmological expansion has been estimated in the framework of Einstein-Strauss model [19]. More investigation is needed in this respect for a definite conclusion. It is also interesting to examine the influence of other matters of the universe to the system. An investigation has been undertaken in this direction.

\end{document}